\documentclass[5p,11pt]{elsarticle}
\usepackage{amsfonts,color}
\usepackage{amsmath}
\usepackage{amssymb}
\usepackage{graphicx,dsfont}
\journal{Physics Letter B}

\begin{document}
\begin{frontmatter}
\title{Number conserving particle-hole RPA for superfluid nuclei}

\author[1]{J.~Dukelsky}
\ead{dukelsky@iem.cfmac.csic.es}

\author[2,3,4]{J.E.~Garc\'{\i}a-Ramos\corref{cor1}}
\ead{enrique.ramos@dfaie.uhu.es}

\author[5,3]{J.M.~Arias}
\ead{ariasc@us.es}

\author[6,3]{P.~P\'erez-Fern\'andez}
\ead{pedropf@us.es}

\author[7,8]{P.~Schuck}
\ead{schuck@ipno.in2p3.fr}

\address[1]{Instituto de Estructura de la Materia, CSIC, Serrano 123, 28006 Madrid, Spain}

\address[2]{Departamento de  Ciencias Integradas y Centro de Estudios
  Avanzados en F\'isica, Matem\'atica y Computaci\'on,\\ Universidad de Huelva,
  21071 Huelva, Spain}

\address[3]{Instituto Carlos I de F\'{\i}sica Te\'orica y Computacional,
  Universidad de Granada, Fuentenueva s/n, 18071 Granada, Spain}

\address[4]{Unidad Asociada de la Universidad de Huelva al IEM (CSIC), Madrid,
  Spain}

\address[5]{Departamento de F\'{\i}sica At\'omica, Molecular y Nuclear,
  Facultad de F\'{\i}sica, Universidad de Sevilla, Apartado~1065,
  41080 Sevilla, Spain}

\address[6]{Departamento de  F\'{\i}sica Aplicada III, Escuela T\'ecnica
  Superior de Ingenier\'{\i}a, Universidad de Sevilla, Camino de los Descubrimientos, 41092 Sevilla, Spain}

\address[7]{Institut de Physique Nucl\'eaire, Universit\'e de Paris-Sud, CNRS-IN2P3, 15 rue G. Clemenceau, 91406 Orsay Cedex, France}

\address[8]{Universit\'e de Grenoble Alpes, CNRS, LPMMC, 38000 Grenoble, France.}

\cortext[cor1]{Corresponding author}

\begin{abstract}
We present a number conserving particle-hole RPA
theory for collective excitations in the transition from normal to
superfluid nuclei. The method derives from an RPA theory developed long ago
in quantum chemistry using antisymmetric geminal powers, or
equivalently number projected HFB states, as reference
states. We show within a minimal model of pairing plus monopole interactions
that the number conserving particle-hole RPA excitations evolve smoothly across the superfluid phase transition close to the
exact results, contrary to particle-hole RPA in the normal phase and
quasiparticle RPA in the superfluid phase that require a change of basis at
the broken symmetry point. The new formalism can be applied in a straightforward manner to study particle-hole excitations on top of a number projected HFB state.
\end{abstract}


\end{frontmatter}


\section{Introduction}
Excitation spectra of nuclei show a tremendous richness and diversity, going from low energy collective states to giant resonances.
The basic approach for the description of those phenomena is time-depen\-dent mean field with the use of more or less sophisticated energy density functionals (EDFs). In the small amplitude limit, this means the consistent solution of  mean field and RPA equations. For superfluid nuclei this has to be generalized to Hartree-Fock-Bogoliubov (HFB) and quasiparticle RPA (QRPA) equations. Ma\-ny works of this type based on Skyrme or Gogny EDFs as well as relativistic HFB exist in the literature, see for example \cite{Sarri, Tera, Goriely, Colo, Ring, Peru}. The weakness of the HFB approach is that it violates, among other symmetries, particle number conservation. This is no problem in macroscopic superfluid systems. However, in finite nuclei this becomes annoying if there is a spread of several units of particles in the calculations where one desires a sharp number for each nucleus. The restoration of good particle number is technically feasible with modern projection techniques \cite{Doba, Robledo} but becomes rather cumbersome on the level of QRPA where one has to project two-quasiparticle states with a subsequent orthogonalization \cite{Kyotoku}. This might be the reason why only in relatively rare cases it has been applied mainly to $\beta$ and double-$\beta$ decays \cite{Civitarese,Suhonen}. Moreover, number projected QRPA, being a projection of two quasiparticle states, involves a mixture of particle-hole (ph) excitations of the $A$ nucleus, particle-particle excitations of the $A-2$ nucleus and hole-hole of the $A+2$ nucleus. At the end one, therefore, one has to filter out the most interesting excitations of the $A$ system which are then excitations of the ph type. These problems are particularly important for transitional regions since there the particle number fluctuations are strong and the transition from normal to superfluid is rounded due to the finiteness of the system. Crossing the superfluid transition requires a change in the reference state from a HF Slater determinant to a HFB quasiparticle vacuum. The collective ph excitations described by RPA in the former case and by QRPA in the latter yield a sharp transition with a significant kink \cite{Krumlinde}.

Collective states with projections, in particular particle number projection, have been investigated with the Generator Coordinate Method (GCM). However, GCM is generally tailored for large amplitude collective motion and only a few collective coordinates are considered which are obtained from constrained and symmetry projected HFB calculations. RPA theory which is considered in the present work, on the contrary, treats collective motion in the small amplitude limit where collective and non-collective states are generally treated on the same footing. A recent paper with many references to the GCM-PHFB method can be found in \cite{Rodriguez}.
In this situation, we thought it useful to adapt a formalism, known in quantum chemistry since many years \cite{Sangfelt}, to the nuclear physics case. While in quantum chemistry this formalism has had little success due the absence of attractive pairing correlations, for the same reason we do believe that it could be extremely useful in nuclear structure. In chemical physics it was shown \cite{Linderberg} that the so-called antisymmetrized geminal power (AGP) state, which is the same as the well known number projected HFB (PHFB) ground state, is the vacuum of a set of ph operators which kill the PHFB vacuum. This remarkable property can be exploited to construct a particle-hole RPA (phRPA) approach along the same lines as in the standard RPA theory. One adds to the particular set of ph killers which play the role of backward terms, their adjoints as the forward terms with the corresponding amplitudes and establishes with the equation of motion technique the RPA matrix. We call this theory number conserving particle-hole RPA (NCphRPA).
This new particle number conserving technique may be implemented without too much difficulty and effort in the existing QRPA codes to collective modes in superfluid nuclei like, e.g., the chain of Sn isotopes. Since NCphRPA adds correlations on top of standard ph-RPA even at magicity, one could consider all Sn isotopes from A = 100 to A = 132 and beyond with the same approach.

 We will first classify the complete set of ph operators in terms of killers of an arbitrary PHFB state, their adjoints and diagonal operators. We will then proceed to derive the NCphRPA based on the killers and adjoints in complete analogy to the RPA over a HF reference state. Finally, we will test numerically this approach in a minimal model of pairing plus monopole ph interactions.\\

\section{PHFB killers}
Let us start by defining the PHFB wave function of $M$ nucleon pairs as a general pair condensate in the canonical basis where the density matrix is diagonal
\begin{equation}
| PHFB \rangle = \Gamma^{\dagger M}\left | 0\right\rangle \text{,
\ }\Gamma^{\dagger}=\sum_{i=1}^{L} z_{i}a_{i}^{\dagger}a_{\overline{i}}^{\dagger
}, \label{N}%
\end{equation}
where the operator $a_{i}^{\dagger}$ ($a_{\overline{i}}^{\dagger}$) create a particle in the state $i$ ($\overline{i}$).  $\left\{  i, \overline{i}\right\}  $ is a single particle basis with $2 L$ states that has been grouped into $L$ pairs $i$ and $\overline{i}$. The $\overline{i}$ states are conjugate orbitals to the ${i}$ single particle states. These pairs are not necessarily degenerated but have the same occupation probabilities $\langle a_{i}^{\dagger} a_i \rangle=\langle a_{\overline{i}}^{\dagger} a_{\overline{i}}\rangle$. The free parameters $z_{i}$ can eventually be determined variationally as for example in HFB with number projected before variation.
Similar to a HF Slater determinant where it is possible to define a set of killers in terms of which the RPA operators are constructed, PHFB has a set of killers that we classify into particle killers and spin killers.
\begin{itemize}
\item Particle killers:
 \begin{equation}
 C_{ij}=z_{i}a_{i}^{\dagger}a_{j}-z_{j}a_{\overline{j}}^{\dagger}%
a_{\overline{i}},  ~~   \forall ~ i \ne j.
\label{particle}
 \end{equation}

\item Spin killers:
\begin{equation}
\left.
\begin{array}
[c]{c}%
S_{ij}^{+}=z_{i}a_{i}^{\dagger}a_{\overline{j}}+z_{j}a_{j}^{\dagger
}a_{\overline{i}}\\
S_{ij}^{-}=z_{i}a_{\overline{i}}^{\dagger}a_{j}+z_{j}a_{\overline{j}}%
^{\dagger}a_{i}%
\end{array}
\right\} ~ \forall  i > j .
\label{spin}
\end{equation}
\end{itemize}
Note that particle killers move nucleons without changing the spin, while spin killers move nucleons flipping the spin.
It is easy to check that all killers commute with the pair operator (\ref{N}), $[C_{ij},\Gamma^{\dagger}]= [S_{ij}^{+},\Gamma^{\dagger}] =[S_{ij}^{-},\Gamma^{\dagger}]=0$, and therefore, each one of the killers annihilates the $PHFB$ wave function (\ref{N})
\begin{equation}
 C_{ij}| PHFB \rangle = S_{ij}^{+}| PHFB \rangle = S_{ij}^{-}| PHFB \rangle =0 .
 \end{equation}

The set of killers (\ref{particle}-\ref{spin}), their adjoints and the diagonal operators $a_{i}^{\dagger}a_{i}$, $a_{\overline{i}}^{\dagger}a_{i}$, $a_{i}^{\dagger}a_{\overline{i}}$ and $a_{\overline{i}}^{\dagger}a_{\overline{i}}$ form a complete set of $4L^2$ ph operators that generate a $U(2L)$ algebra of particle-hole operators defined by the commutators
\begin{equation}
\left[ a_{\alpha}^{\dagger}a_{\beta},a_{\gamma}^{\dagger}a_{\delta}\right]
=\delta_{\beta\gamma}a_{\alpha}^{\dagger}a_{\delta}-\delta_{\alpha\delta
}a_{\gamma}^{\dagger}a_{\beta},
\label{comm}
\end{equation}
where $\alpha,\beta,\gamma,\delta$ stand for any orbital $j$ and their conjugate, $\overline{j}$. Within this set there are a total of $L(L-1)$ particle killers and the same number of spin killers.  The total number of adjoints that create particle-type or spin-type excitations is $2L(L-1)$. Adding up the killers, the adjoints and the $4L$ diagonals, we obtain the total of $4L^2$ ph operators. The completeness of this set guaranties that any ph operator can be expressed as a linear combination of the above generators
\begin{align}
a_{i}^{\dagger}a_{j}&=\frac{1}{z_{i}^{2}-z_{j}^{2}}\left( z_{i}C_{ij}%
-z_{j}C_{ji}^{\dagger}\right) ,\nonumber\\
a_{\overline{i}}^{\dagger}a_{\overline{j}}&=\frac{-1}{z_{i}^{2}-z_{j}^{2}%
}\left(  z_{i}C_{ji}-z_{j}C_{ij}^{\dagger}\right),\nonumber\\
a_{i}^{\dagger}a_{\overline{j}}&=\frac{1}{z_{i}^{2}-z_{j}^{2}}\left[
z_{i}S_{ij}^{+}-z_{j}\left(  S_{ij}^{-}\right)  ^{\dagger}\right].
\label{Inv}
\end{align}

The denominators that appear in the inversion relations could be the source of numerical instabilities in case of level crossings or for states deep in the Fermi sea with occupations close to 1. In the former case, the pair of approaching levels could be considered as effectively degenerated and redefine de killers taking into account this degeneracy. In the latter case, it might be possible to consider these states as an inert core and exclude them from the numerical calculation.  
On the other hand, degeneracies due to the conservation of symmetries like the rotational symmetry (spherical nuclei) can be explicitly taken into account in the definition of the killers.

Using the commutation algebra of the ph operators (\ref{comm}) it is easy to derive the commutation relations between killers
\begin{align}
  [C_{ij},S_{pq}^{+}]&=z_pS_{iq}^{+} \delta_{jp}+z_qS_{ip}^{+}\delta_{jq},\nonumber\\
  [C_{ij},S_{pq}^{-}]&=-z_qS_{pj}^{-}\delta_{iq}-z_pS_{qj}^{-}\delta_{ip}, \nonumber\\
  [C_{ij},C_{pq}]&=z_pC_{iq}\delta_{jp}-z_qC_{pj}\delta_{iq}, \nonumber\\
  [S_{ij}^{+},S_{pq}^{-}]&= z_p(C_{iq}\delta_{jp}+C_{jq}\delta_{ip})\nonumber\\
  &+z_q(C_{jp}\delta_{iq}+C_{ip}\delta_{jq}),\nonumber\\
  [S_{ij}^{+},S_{pq}^{+}]&=0,  [S_{ij}^{-},S_{pq}^{-}]=0.
  \label{comm-killers}
\end{align}

The complete set of commutators including the adjoints and diagonals follows in a similar manner.

\section{Number conserving particle-hole RPA}
We will derive the NCphRPA equations following closely the equation of motion method of \cite{Rowe, Ring-Schuck} assuming a PHFB pair condensate (\ref{N}) as the reference state. In order to ease the formalism we restrict to particle excitations, but the generalization to spin excitations is straightforward. In fact, if the PHFB state is axially symmetric, excitations with a given $k$ value of the $z$ component of the total angular momentum could be constructed as a mixture of particle and spin excitations with the same $k$.

As usual, we assume that excited states $\left\vert \mu\right\rangle =Q_{\mu
}^{\dagger}\left\vert 0\right\rangle $ are generated by collective RPA
operators

\begin{figure}[tbp]
\centering
\includegraphics[width=0.7\columnwidth]{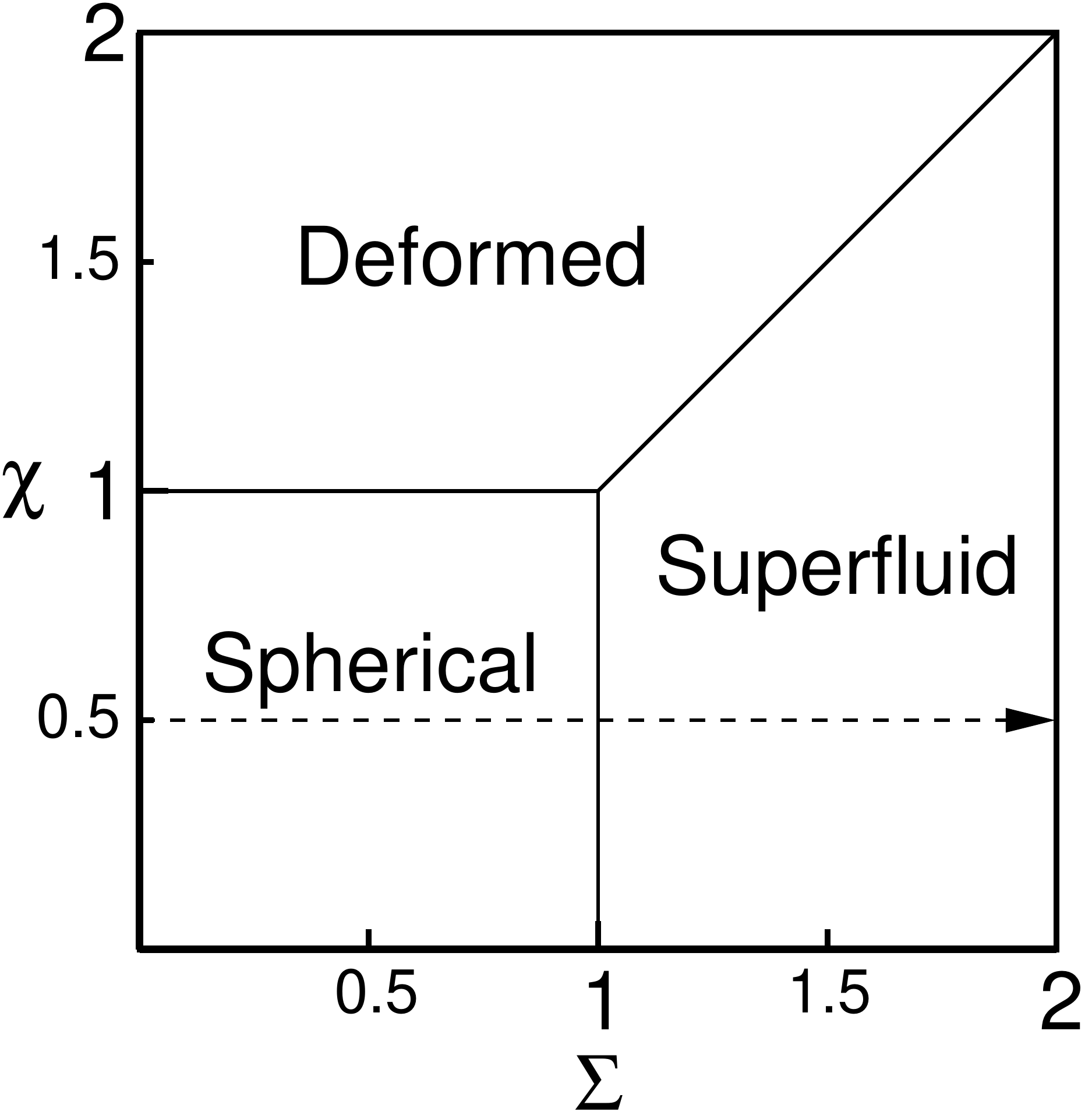}

\caption{Phase diagram of the Agassi Hamiltonian (\ref{Agassi}). The dotted line with $\chi=1/2$ will be used to benchmark the NCphRPA.}
\label{fig:phasediagram}
\end{figure}

\begin{equation}
Q_{\mu}^{\dagger}=\sum_{i > j}X_{ij}^{\mu}D_{ij}^{\dagger}-Y_{ij}^{\mu
}D_{ij}\text{, \ }D_{ij}=\frac{C_{ij}}{\sqrt{(z^2_{i}-z^2_{j})(n_{i}-n_{j})}},
\label{Q}
\end{equation}
where the square root is a normalization factor and $n_{i}=\langle
a_{i}^{\dagger}a_{i}\rangle $ is the occupation probability in the PHFB
state. With this definition, the backward term annihilates the PHFB state in complete analogy to the RPA\ based on HF.

Moreover, the RPA assumes a killing condition
\begin{equation}
Q_{\mu}\left\vert RPA\right\rangle =0,
\end{equation}
for the correlated RPA ground state which later will be relaxed replacing it by
the PHFB reference state.

We can now check that the RPA operators (\ref{Q}) are properly normalized in
the PHFB state
\begin{equation}
\left\langle PHFB\right\vert [Q_{\nu},Q_{\mu}^{\dagger}]\left\vert PHFB\right\rangle
=\sum_{i > j}X_{ij}^{\nu}X_{ij}^{\mu}-Y_{ij}^{\nu}Y_{ij}^{\mu}=\delta
_{\mu\nu}.
\end{equation}
As well as
\begin{equation}
\left\langle PHFB\right\vert [Q_{\nu},Q_{\mu}]\left\vert PHFB\right\rangle
=\sum_{i > j}X_{ij}^{\nu}Y_{ij}^{\mu}-X_{ij}^{\nu}Y_{ij}^{\mu}=0.
\end{equation}
Together with the closure relations
\begin{equation}
\sum_{\mu}\left(  X_{ij}^{\mu}X_{kl}^{\mu}-Y_{ij}^{\mu}Y_{kl}^{\mu}\right)
=\delta_{ik}\delta_{jl},%
\end{equation}
\begin{equation}
\sum_{\mu}\left(  Y_{ij}^{\mu}X_{kl}^{\mu}-X_{ij}^{\mu}Y_{kl}^{\mu}\right)  =0,
\end{equation}
it allows us to invert the RPA operators%
\begin{equation}
D_{ij}=\sum_{\mu}X_{ij}^{\mu} Q_{\mu}+Y_{ij}^{\mu}
Q_{\nu}^{\dagger},
\label{D}
\end{equation}

\begin{equation}
D_{ij}^{\dagger}=\sum_{\mu}X_{ij}^{\mu} Q_{\mu}^{\dagger}%
+Y_{ij}^{\mu} Q_{\nu}.
\label{DD}
\end{equation}

The equation of motion method leads to the RPA equation
\begin{equation}
\left(
\begin{array}{cc}
A & B \\
-B^{\ast} & -A^{\ast}%
\end{array}
\right) \left(
\begin{array}{c}
X \\
Y%
\end{array}
\right) =E_{\mu}\left(
\begin{array}{c}
X \\
Y%
\end{array}
\right),
\end{equation}
with
\begin{equation}
A_{ijkl}=\left\langle PHFB\right\vert \left[ D_{ij},\left[ H,D_{kl}^{\dagger }%
\right] \right] \left\vert PHFB\right\rangle ,\label{A}
\end{equation}

\begin{equation}
B_{ijkl}=-\left\langle PHFB\right\vert \left[ D_{ij}^{\dagger },\left[ H,D_{kl}^{\dagger }\right] %
\right] \left\vert PHFB\right\rangle . \label{B}
\end{equation}

\begin{figure}[tbp]
\centering
\includegraphics[width=.7\columnwidth]{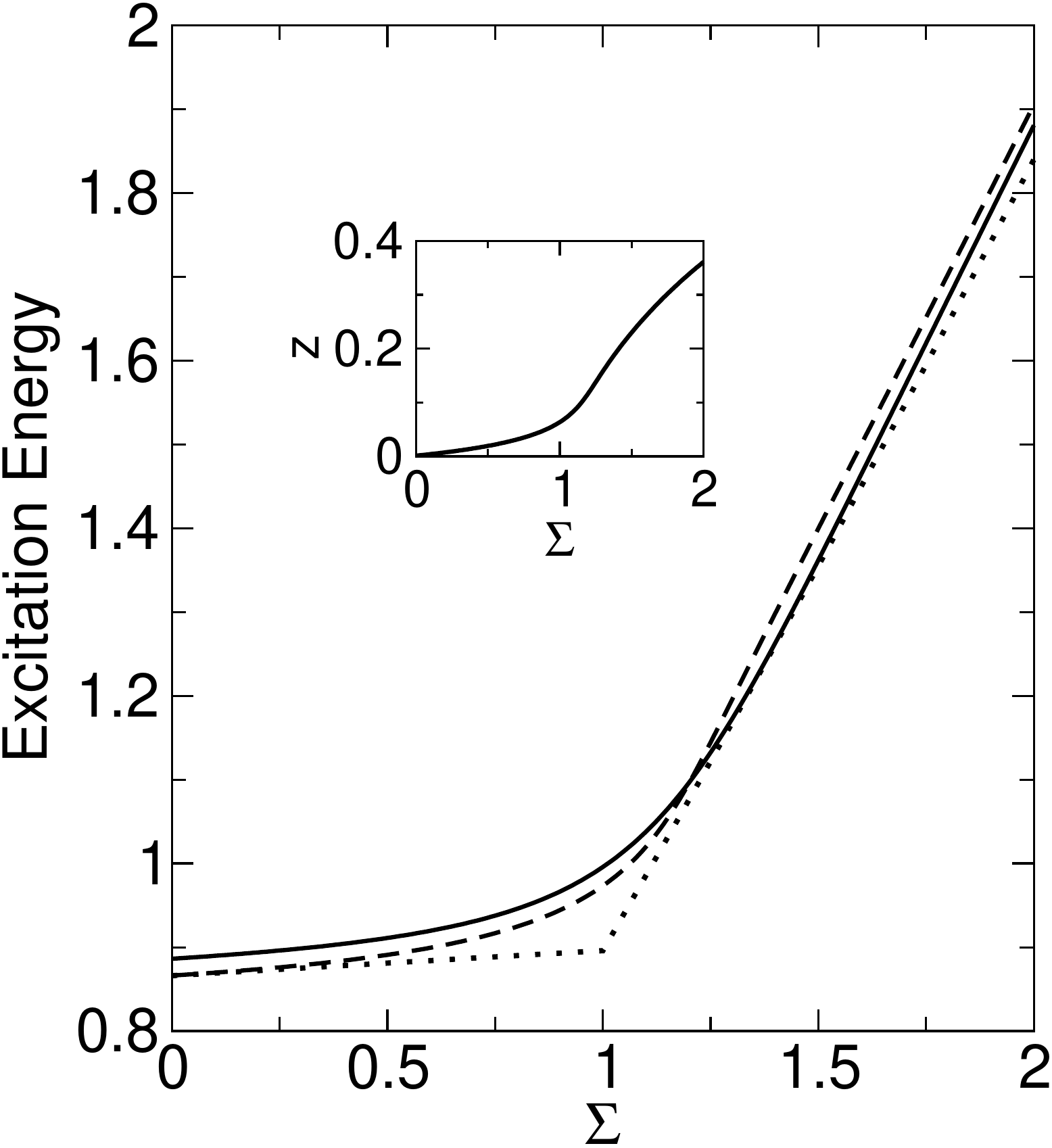}
\caption{First excited particle-hole state of negative parity for a system with degeneracy $j=10$ at half filling. For $\protect%
\chi =0.5$,  as a function of $\Sigma$ the system crosses the transition from spherical to superfluid at $\Sigma=1$. The continuous line represents the exact excited state energy,
the dashed line NCphRPA, and the doted line the phRPA in spherical phase and the QRPA in the superfluid phase. The inset shows the value of $z=\frac{z_+}{z_-}$ along the transition}
\label{fig:Energy}
\end{figure}

Given a transition operator $T$, its transition matrix element $T_{\mu}= \left\langle 0 \right\vert T \left\vert \mu\right\rangle $  can be evaluated as
\begin{eqnarray}
T_{\mu}&=&\sum_{i > j}X_{ij}^{\mu}\left\langle PHFB\right\vert \left[
T,D_{ij}^{\dagger}\right] \left\vert PHFB\right\rangle \nonumber \\
&&-Y_{ij}^{\mu}\left\langle PHFB\right\vert \left[ T,D_{ij}\right] \left\vert
PHFB\right\rangle .
\end{eqnarray}

In NCphRPA we evaluate expectation values of the $A$ and $B$ matrices and the transition matrices in the PHFB ground state in a similar way as the standard RPA replaces the correlated vacuum defined by the killing condition by the mean-field reference state, either a HF Slater determinant or a quasiparticle vacuum. However, the inversion relations (\ref{Inv}) and (\ref{D}-\ref{DD}) would allow us to go beyond NCphRPA towards self-consistent RPA \cite{SCRPA} evaluating most of the expectation values in the true RPA vacuum using the killing condition. \\

\section{Benchmark with the Agassi model}
As a minimal model to benchmark the NCphRPA approximation, we will consider
a two-level model with pairing and particle-hole monopole interactions. The
model was introduced  by Agassi \cite{Agassi} as a tool to test many-body theories that deal with the interplay between ph and superfluid correlations. It has been studied at the mean-field level by Agassi himself \cite{Agassi} and later by Davis and Heiss \cite{Davis}. They derived the phase diagram of the model and the collective excitations in each of the phases by means of HFB and, phRPA and QRPA approximations respectively. More advanced many-body methods like the merging of Coupled Cluster with symmetry restored HFB   theory \cite{Scuseria} revived the model as an excellent test-bed. Recently, Garc\'ia-Ramos {\it et al} \cite{Ramos}  generalized the model by the introduction of new interaction terms that give rise to an extremely rich phase diagram.
We will use here the original Agassi model described by a Hamiltonian which is a superposition
of the Lipkin model and the two-level pairing model
\begin{equation}
H = J_0
-\frac{\Sigma}{(2j-1)}\sum_{\sigma\sigma^{\prime}}A^{\dagger}_{\sigma}A_{\sigma^{\prime}} -\frac{%
\chi}{2(2j-1)}[ J^2_+ + J_-^2 ] , \label{Agassi}
\end{equation}
where $\sigma = \pm 1$ labels each of the two single particle levels with degeneracy $2 j$, and $\Sigma$ and $\chi$ are suitably scaled coupling parameters for pairing and ph interactions. The pair creations operators are
\begin{eqnarray}
  &~& A^{\dagger}_\sigma=\sum_{m=1}^ja^{\dagger}_{\sigma,m}a^{\dagger}_{\sigma,-m}, \\
  &~& A_{0}^{\dagger}  =\sum_{m=1}^{j}\left( a_{-1m}^{\dagger}a_{1,-m}^{\dagger
}-a_{-1-m}^{\dagger}a_{1,m}^{\dagger}\right)
\end{eqnarray}
and the ph operators are
\begin{equation}
J_+= \sum_{m=-j}^j a^{\dagger}_{1m}a_{-1m} = (J_-)^{\dagger},
 \end{equation}
\begin{equation}
J_{0}=\frac{1}{2}\sum_{m=-j}^{j}\left( a_{1m}^{\dagger}a_{1m}-a_{-1m}^{\dagger }a_{-1m}\right).
 \end{equation}

\begin{figure}[tbp]
\centering \includegraphics[width=.7\columnwidth]{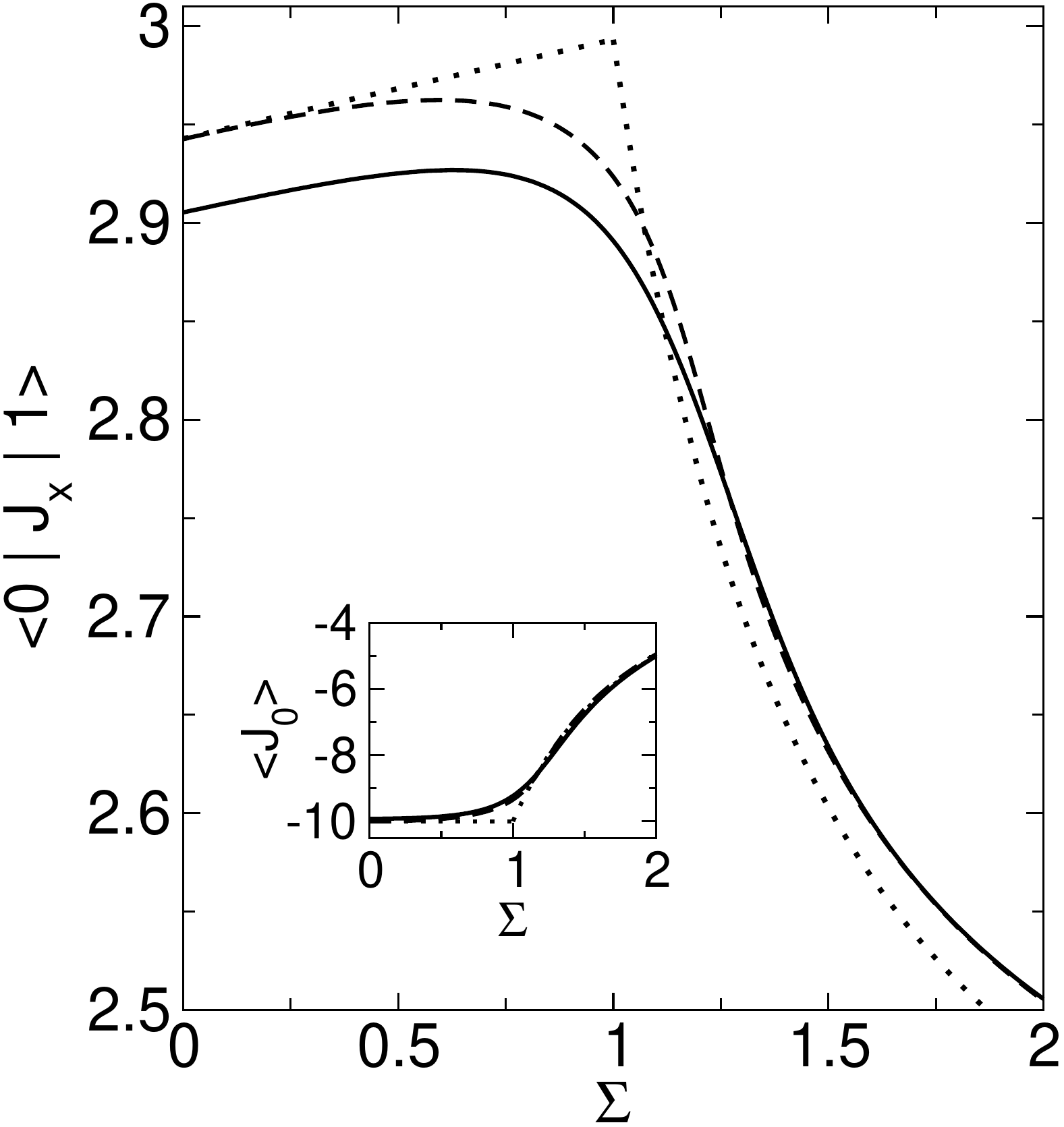}
\caption{Transition matrix element of $J_{x}$ between the
excited state and the GS. The inset shows the ground state expectation value of the operator $J_0$. The convention for the exact, NCphRPA, and phRPA and QRPA is the same as in Fig.\ \ref{fig:Energy}.   }
\label{fig:Transition}
\end{figure}

Fig.\ \ref{fig:phasediagram} shows the phase diagram of the Agassi model at half filling \cite{Davis} with a normal (spherical) phase for $\chi<1$ and $\Sigma<1$, a parity broken (deformed) phase for $\chi>1$ and $\chi>\Sigma$, and a superfluid phase for $\Sigma>1$ and $\Sigma>\chi$, for details see \cite{Davis, Ramos}. Interestingly, the model has no stable phase with both symmetries broken.
We will test and compare the different RPA approximations along the horizontal line $\chi=\frac{1}{2}$ which contains sizeable ph correlations. As a function of $\Sigma$, it crosses the transition from normal to superfluid requiring a change of the reference state from a Slater determinant (HF) to a quasiparticle vacuum (BCS).
Since the line $\chi=\frac{1}{2}$ runs across the normal and superfluid regions that preserve parity, we use the following PHFB ansatz at half filling
\begin{equation}
\left\vert PHFB\right\rangle =\Gamma^{\dagger j}\left\vert 0\right\rangle ,%
\text{ \ }\Gamma^{\dagger}=z_{-}A_{-1}^{\dagger}+z_{+}A_{1}^{\dagger} ,
\label{PC}
\end{equation}
where the structure parameters $z_{\pm}$ can be obtained by a straightforward minimization of the energy or by number projected HFB calculation before variation. In our case, we expand (\ref{PC}) using a binomial
\begin{equation}
\left\vert PHFB\right\rangle =\sum_{l=0}^{j}\frac{j!}{\left(  j-l\right)
!l!}z_{+}^{l}z_{-}^{\left(  j-l\right)  }A_{+1}^{\dagger l}A_{-1}^{\dagger
\left(  j-l\right)  }\left\vert 0\right\rangle .
\label{Bin}
\end{equation}

Note that for $z_+=0$ we recover the HF Slater determinant with all particles filling the lower level.

PHFB improves over the mean field used to construct the phase diagram taking into account some of the correlations even in the normal phase.

Due to the simplicity of the model, we are left with a couple of ph operators $J_+$  and $J_-$, in terms of which the killer of the pair condensate (\ref{PC}) is
\begin{equation}
C=z_{-} J_{-} - z_{+} J_{+} .
 \end{equation}

It can be easily checked that $[C,\Gamma^{\dagger}]=0$ and therefore, $C$ annihilates the pair condensate (\ref{PC}).
Following the general derivation of the NCphRPA, the ph RPA operator is
\begin{equation}
  Q^{\dagger}=XD^{\dagger}-YD,~~D=\frac{C}{\sqrt{2\langle J_0 \rangle (z_{+}^2-z_{-}^2)}} .
\end{equation}

The RPA matrices $A$ and $B$ as defined in (\ref{A}) and (\ref{B}) can be obtained from the double commutators
\begin{align}
&\left[  C,\left[  H,C^{\dagger}\right] \right]     =-2\varepsilon\left(
z_{-}^{2}+z_{+}^{2}\right)  J^{0}+2\frac{\chi}{2j-1}\left[  z_{-}^{2}%
J^{-2}\right.  \nonumber\\
& \left.  +z_{+}^{2}J^{+2}+z_{+}z_{-}\left(  J^{+}J^{-}+J^{-}J^{+}%
-4J^{02}\right)  \right]  \nonumber\\
& -2\frac{\Sigma}{2j-1}\left(  z_{-}-z_{+}\right)  ^{2}A_{0}^{\dagger}%
A_{0}\nonumber\\
& +2\frac{\Sigma}{2j-1}\left(  z_{-}-z_{+}\right)  ^{2}\left(  A_{1}^{\dagger
}A_{1}+A_{-1}^{\dagger}A_{-1}\right)  \nonumber\\
& +4\frac{\Sigma}{2j-1}\left(  z_{-}-z_{+}\right)  \left(  z_{-}%
A_{-1}^{\dagger}A_{1}-z_{+}A_{1}^{\dagger}A_{-1}\right)
\end{align}
and
\begin{align}
&\left[  C^{\dagger},\left[  H,C^{\dagger}\right]  \right]     =4\varepsilon
z_{+}z_{-}J^{0}-2\frac{\chi}{2j-1}z_{+}z_{-}\left[  \left(  J^{+2}%
+J^{-2}\right)  \right. \nonumber \\
& \left.  +\left(  z_{+}^{2}+z_{-}^{2}\right)  \left(  J^{-}J^{+}+J^{+}%
J^{-}-4J^{02}\right)  \right] \nonumber \\
& -\frac{\Sigma}{2j-1}\left(  z_{-}-z_{+}\right)  z_{-}\left[  2A_{0}%
^{\dagger}A_{0}\right. \nonumber  \\
& \left.  -2\left(  A_{1}^{+}+A_{-1}^{+}\right)  A_{-1}-2A_{1}^{\dagger
}\left(  A_{1}+A_{-1}\right)  \right] \nonumber \\
& +\frac{\Sigma}{2j-1}\left(  z_{-}-z_{+}\right)  z _{+}\left[ 2A_{0}%
^{\dagger}A_{0}\right.  \nonumber\\
& \left.  -2\left(  A_{1}^{\dagger}+A_{-1}^{\dagger}\right)  A_{1}%
-2A_{-1}^{\dagger}\left(  A_{1}+A_{-1}\right)  \right] ,
\end{align}
by taking expectation values in the  PHFB state (\ref{Bin}).

It can be easily checked that NCphRPA reduces to phRPA if we replace the PHFB reference state (\ref{Bin}) by the Slater determinant obtained in the limit $z_+=0$ with all particles occupying the lower level.

In Fig.\ \ref{fig:Energy} we show the excitation energy\\ $\omega=\sqrt{A^2-B^2}$ along the line of $\chi=0.5$, as a function of $\Sigma$. It is compared with the phRPA for $0<\Sigma<1$ in the spherical regions and with QRPA for $\Sigma>1$ in the superfluid region \cite{RPAS}. We can see a clear improvement of
NCphRPA over both RPA's, specially around the phase transition which NCphRPA crosses smoothly, while in the conventional RPA one has to change from a number conserving HF basis to a number non-conserving quasiparticle vacuum with a consequent kink. This improvement could be traced back to the structure of the PHFB reference state shown in the inset. This parameter $z=\frac{z_+}{z_-}$ displays a smooth behavior across the transition due to the inclusion of pair correlations and number fluctuations absent at the mean-field level.

In order to test the RPA wave function we evaluate the transition matrix element of the operator $J_x = \frac{1}{2} ( J_+ + J_- )$.
\begin{eqnarray}
&&\left\langle PHFB \right\vert [ J_x , Q^{\dagger} ] \left\vert PHFB\right\rangle
= \nonumber\\
&&\left( X+Y\right) \left( z_{+}+z_{-}\right) \sqrt{\frac{\left\langle
J_{0}\right\rangle }{2\left( z_{+}^{2}-z_{-}^{2}\right) }} .
\label{tran}
\end{eqnarray}

In Fig.\ \ref{fig:Transition} we show this transition matrix
element. Again we see a clear improvement of the present approach over
the conventional RPA's \cite{RPAS}, mainly in the transitional region. The properties of the excited state as described by the NCphRPA reflect the net improvement seen in the ground state at the level of PHFB as seen in the inset of Fig.\ \ref{fig:Energy} as well as in the inset of Fig.\ \ref{fig:Transition} for the expectation value of $J_0$.
\section{Conclusions}
In this work we have introduced the NCphRPA following a similar approach presented long ago in quantum chemistry. NCphRPA starts with a linear transformation of the one particle operators such that they can be classified into killers of a PHFB state, their adjoints, and diagonals. Based on this new one particle operators, NCphRPA is derived in a standard way using the equations of motion method. The formalism is a straightforward extension of phRPA based on a HF determinant as a reference state, and it requires the sole information of the one and two-body density matrices in the optimized PHFB state. In this regard, it is much simpler than the number projected QRPA that requires the projection of a complete set of two quasiparticle states and the subsequent re-orthonormalization.
The general formalism  of NCphRPA developed and tested here is ready to be used with density-dependent forces treated with PHFB to describe collective motion in transitional and superfluid nuclei. Even for magic nuclei, NCphRPA with a PHFB reference state accounting for some of the pair fluctuations, would give non-trivial corrections to phRPA based on a Slater determinant. Therefore, NCphRPA can be used with benefit throughout the mass table. These kinds of corrections to phRPA can be seen in figures 2 and 3 for $0<\Sigma<1$. We want also to stress that the transition from non-superfluid to superfluid nuclei is always completely smooth since the reference state (PHFB) does not show any abrupt change in the crossover. The fact that the NCphRPA uses the equation of motion method to derive the RPA equations allows for an improved treatment of the reference in the direction of a number conserving self-consistent RPA, by using the RPA killing condition and the inversion of the RPA operators.
Before closing we would like to emphasize the ph character of the NCphRPA. Within this formalism particle-particle correlation are only described at the level of the PHFB reference state. However, particle-particle and hole-hole fluctuations could be taken into account in a second RPA based on quadratic killers and quadratic adjoints.

\section{ Acknowledgements}
This work has been partially supported by the Consejer\'{\i}a de Econom\'{\i}a, Conocimiento, Empresas y Universidad de la Junta de Andaluc\'{\i}a (Spain) under Groups FQM-160 and FQM-370 and by European Regional Development Fund (ERDF), ref.\  SOMM17/\\6105/UGR.
We acknowledge financial support from the Spanish Ministerio de Ciencia, Innovaci\'on y Universidades and the ERDF under Projects No.\ FIS2015-63770-P, FIS2017-88410-P and PGC2018-094180-B-I00. Resources supporting this work were provided by the CEAFMC and Universidad de Huelva High Performance Computer (HPC@UHU) funded by \\FEDER/MINECO project UNHU-15CE-2848.


\begin{thebibliography}{9}

\bibitem{Sarri} P. Sarriguren, E. M. de Guerra, and R. Nojarov, Phys. Rev. C \textbf{54}, 690 (1996).

\bibitem{Tera} J. Terasaki, and J. Engel,  Phys. Rev. C \textbf{82}, 034326 (2010).

\bibitem{Goriely} S. Goriely, S. Hilaire, S. Peru, M. Martini, I. Deloncle, F.
Lechaftois, Phys. Rev. C \textbf{94}, 044306 (2016).


\bibitem{Colo} E. Y\"{u}ksel, G. Col\`{o}, E. Khan, Y. F. Niu, Phys. Rev. C \textbf{97}, 064308 (2018).

\bibitem{Ring} D. P. Arteaga, and P. Ring,  Phys. Rev. C \textbf{77} 034317 (2008).

\bibitem{Peru} S. P\'{e}ru, and M. Martini, Eur. Phys. J. A \textbf{50}, 88 (2014).

\bibitem{Doba} J. Dobaczewski, M. V. Stoitsov, W. Nazarewicz, and P.-G. Reinhard, Phys. Rev. C \textbf{76}, 054315 (2007).

\bibitem{Robledo}  J. A. Sheikh, J. Dobaczewski, P. Ring, L. M. Robledo, C. Yannouleas, arXiv:1901.06992.

\bibitem{Kyotoku} M. Kyotoku, K. W. Schmid, F. Gr\"{u}mmer, and A. Faessler,
Phys. Rev. C \textbf{41}, 284 (1990).

\bibitem{Civitarese} O. Civitarese, A. Faessler, J. Suhonen, and X. R. Wu,
Nucl. Phys. A \textbf{524}, 404 (1991).

\bibitem{Suhonen} J. Suhonen, J. Phys. G \textbf{19}, 139 (1993).

\bibitem{Krumlinde} J. Bang, and J. Krumlinde, Nucl. Phys. A \textbf{141}, 1 (1970).


\bibitem{Rodriguez} L. M. Robledo, T. R. Rodr\'{i}guez, R. R. Rodr\'{i}guez-Guzm\'{a}n, J. Phys. G \textbf{46}, 013001 (2019).


\bibitem{Sangfelt} E. Sangfelt, R. Roy Chowduri, B. Weiner, and Y. \"{O}hrn,
J. Chem. Phys. \textbf{88}, 4523 (1987).

\bibitem{Linderberg} J. Linderberg, and Y. \"{O}hrn, Int. J. Quantum Chem.,
\textbf{12}, 161 (1977).

\bibitem{Rowe} D. J. Rowe, Rev. Mod. Phys. \textbf{40}, 153 (1968).

\bibitem{Ring-Schuck} P. Ring, and P. Schuck, \textit{{The Nuclear Many-Body
Problem}, Springer (1980). }

\bibitem{SCRPA} J. Dukelsky, P. Schuck, Nucl. Phys. A \textbf{512}, 466 (1990); J. Dukelsky, G. Roepke, P. Schuck, Nucl. Phys. A \textbf{628}, 17 (1998).


\bibitem{Agassi} D. Agassi, Nucl. Phys. A \textbf{116}, 49 (1968).


\bibitem{Davis} E. D. Davis, and W. D. Heiss, J. Phys. G: Nucl. Phys. \textbf{12}, 805 (1986).

\bibitem{Scuseria} M. R. Hermes, J. Dukelsky, and G. E. Scuseria, Phys. Rev. C \textbf{95}, 064306 (2017).

\bibitem{Ramos} J. E. Garc\'{i}a-Ramos, J. Dukelsky, P. P\'{e}rez-Fern\'andez, and J. M. Arias, Phys. Rev. C \textbf{97}, 054303 (2018).

\bibitem{RPAS} The solution of the phRPA and QRPA for the Agassi model can be found in \cite{Davis}.



\end{thebibliography}
\end{document}